\documentclass[conference]{IEEEtran}
\usepackage{cite}
\usepackage{amsmath,amssymb,amsfonts}
\usepackage{algorithmic}
\usepackage{graphicx}
\usepackage{textcomp}
\usepackage{xcolor}

\usepackage{physics}
\usepackage{qcircuit}
\usepackage[capitalise]{cleveref}

\graphicspath{{./Figures/}}

\usepackage{graphicx}
\usepackage{subcaption}
\usepackage[textfont=normalfont,singlelinecheck=off,justification=raggedright]{caption}

\newcommand{\CZ}{$\mathsf{CZ}$ }

\def\BibTeX{{\rm B\kern-.05em{\sc i\kern-.025em b}\kern-.08em
    T\kern-.1667em\lower.7ex\hbox{E}\kern-.125emX}}

\begin{document}

\title{Designing high-fidelity multi-qubit gates for semiconductor quantum dots through deep reinforcement learning}

\author{\IEEEauthorblockN{Sahar Daraeizadeh}
\IEEEauthorblockA{\textit{Intel Labs.} \\
\textit{Intel Corporation}\\
Hillsboro, USA\\
sahar.daraeizadeh@intel.com}
\and
\IEEEauthorblockN{Shavindra P. Premaratne}
\IEEEauthorblockA{\textit{Intel Labs.} \\
\textit{Intel Corporation}\\
Hillsboro, USA\\
Shavindra.Premaratne@intel.com}
\and
\IEEEauthorblockN{A. Y. Matsuura}
\IEEEauthorblockA{\textit{Intel Labs.} \\
\textit{Intel Corporation}\\
Hillsboro, USA\\
anne.y.matsuura@intel.com}
}

\maketitle

\begin{abstract}
In this paper, we present a machine learning framework to design high-fidelity multi-qubit gates for quantum processors based on quantum dots in silicon, with qubits encoded in the spin of single electrons. In this hardware architecture, the control landscape is vast and complex, so we use the deep reinforcement learning method to design optimal control pulses to achieve high fidelity multi-qubit gates. In our learning model, a simulator models the physical system of quantum dots and performs the time evolution of the system, and a deep neural network serves as the function approximator to learn the control policy. We evolve the Hamiltonian in the full state-space of the system, and enforce realistic constraints to ensure experimental feasibility.
\end{abstract}


\section{Introduction}
Over the past 20 years, numerous proposals have been put forward to design semiconductor quantum-dots-based qubits. While quantum dot qubits were always attractive for scalability due to their small form factor, they are also particularly sensitive to magnetic and charge noise \cite{Kuhlman2013} from their environment. Various techniques have been used to help mitigate these noise sources, such as using purified isotopes with zero nuclear spin (like Silicon-28) to reduce magnetic noise \cite{Veldhorst2015} and operating qubits in “sweet spots” that are less susceptible to electrical noise \cite{Shim2016}. Recent research shows higher temperature (1 K) operation of silicon quantum dot qubits \cite{Petit2020}. These advances make quantum dot qubits a compelling system to study for scalable quantum computing.

Quantum-dots-based qubits can be broadly categorized into qubits based on spin states \cite{Loss1998, Hanson2007, Zwaneburg2013}, charge states \cite{Petersson2010, Hayashi2003, Fujisawa2004}, or spin-charge hybrid states \cite{Shi2012, Kim2014, Koh2012}. Spin-based qubit types  can be further distinguished as single-electron spin qubits \cite{Koppens2006, Nowack2007}, singlet-triplet qubits \cite{Petta2005, Foletti2009}, and exchange-only spin qubits \cite{Medford2013a, Medford2013b}. Single electron spin qubits are conceptually one of the simplest types of qubits, and have garnered great interest due to their long lifetimes. Coherent single-qubit operations in such systems are typically performed using electron spin resonance \cite{Koppens2006} or electric-dipole-induced spin resonance \cite{Golovach2006, Kawakami2014}. Two-qubit operations in such systems are typically performed using precise pulsing/activation of the effective exchange interaction between two electrons \cite{Petta2005} or using state selective single-qubit driving \cite{Watson2018}. Depending on the method of control of the effective exchange coupling between the two electrons, different operations such as $\mathsf{\sqrt{SWAP}}$ \cite{Martins2016, Brunner2011}, and \CZ \cite{Veldhorst2015, Watson2018} can be realized.

Optimal control techniques and machine learning approaches have demonstrated great success in design and optimization of multi-qubit gates in various quantum technologies \cite{Chunlin_Chen_2014, google_trpo, daraeizadeh2019, allen_2017, Zahedinejad_2015}, \cite{Zahedinejad_2016, Goerz_2017, Machnes_2011, Spiteri_2018, Khaneja, Rach_2015}. However, the machine learning methods to design and control multi-qubit gates in quantum-dot qubit systems are yet to be investigated. In this paper, we model the multi-qubit gate design problem within a deep Reinforcement Learning (RL) framework \cite{LapanBook}, and explore different learning algorithms based on Temporal Difference (TD) and Policy Gradient methods \cite{SuttonBook} for designing a \CZ gate for quantum-dot qubit systems. The learning frameworks presented here can be adapted for other quantum gate physical realizations as well.

\section{Quantum-dot Gate Design Framework}

\subsection{Hamiltonian Model}

We consider a coupled quantum-dot qubit system with each dot having occupancy up to two electrons. We further constrain each quantum dot to the four lowest energy levels: empty ($0$), single-electron spin-down ($\downarrow$), single-electron spin-up ($\uparrow$), two-electrons in the singlet state ($S$). For simplicity in modeling, we ignore the valley degree of freedom and write the Hamiltonian for the full system as
\begin{equation} \label{hamiltonian} 
\mathcal{H}= \mathcal{H}_\epsilon + \mathcal{H}_Z + \mathcal{H}_U + \mathcal{H}_T,
\end{equation} 
where $\mathcal{H}_\epsilon$, $\mathcal{H}_Z$, $\mathcal{H}_U$, $\mathcal{H}_T$ each correspond to the summations of on-site energies, Zeeman terms, Coulomb repulsion terms (Hubbard interaction energies), and tunnel couplings, respectively \cite{Yang2011, Veldhorst2015}. The on-site energy offset $\epsilon$ of each dot, and the tunnel coupling $t$ between the two dots are controlled through the corresponding plunger and barrier gate voltages, respectively. We construct the Hamiltonian of dimension $16 \times 16$ for a two-dot system following the convention in \cite{mcclean2017openfermion}.

\subsection{Simulation of Quantum System Dynamics}
For a quantum system with a small number of qubits, one can use classical compute resources to simulate the dynamics by time evolving the quantum system. Given the Hamiltonian of a quantum system, the time evolution of the system is governed by the time-dependent Schr\"{o}dinger equation
\begin{equation}
\ket{\Psi(t)} = \exp\left[\frac{-i H(t)\, t}{\hbar} \right] \ket{\Psi(t_0)},
\end{equation}
\noindent where $H(t)$ is the Hamiltonian of the system at time $t$, $\ket{\Psi(t_0)}$ is the initial state of the system, and $\ket{\Psi(t)}$ is the state at time $t$. We use natural units and set $\hbar  = 1$. Considering that the Hamiltonian of the system is time-independent in very small Trotter time steps \cite{Steck2020, Yung_guzik14}, one can solve the time evolution of the system to realize the unitary transformation $U$ during the gate operation.
\begin{equation}
\label{Unitary_exp} 
U = \prod_0^{\tau_\mathrm{total}} \exp \left[ -i H(t_j)\, dt_j \right]
\end{equation}
where $\tau_\mathrm{total}$ is the total gate duration, and the chronological order preserving product is taken over infinitesimal timesteps $t_j$ in the range $(0, \tau_\mathrm{total})$. In our simulations,  first the unitary transformation $U_{16 \times 16}$ is calculated by time evolution of the full Hamiltonian of the two-dot system. Then it is projected to the two-qubit computational subspace considering only matrix elements $\{5,6,9,10\}$ corresponding to states $\{\downarrow\downarrow,\downarrow\uparrow, \uparrow\downarrow, \uparrow\uparrow\}$. Finally we perform phase compensation by removing single-qubit rotations collected by each qubit \cite{daraeizadeh2019}.

In quantum gate design, typically the Hamiltonian control parameters are adjusted by maximizing the gate fidelity such that a target unitary transformation is realized. The gate fidelity $\mathcal{F}$ is defined as \cite{Pedersen_2007}:
\begin{equation} \label{gate_fidelity_ML} 
\mathcal{F}=\frac{{\mathrm{Tr} \left(U^\dagger_\mathrm{final} U_\mathrm{final}\right)} + {\left|{\mathrm{Tr} \left(U^\dagger_\mathrm{target} U_\mathrm{final}\right)}\right|}^2}{d(d+1)}  
\end{equation} 
here $U_\mathrm{target}$ is the ideal target unitary transformation, and  $d=2^{\mathrm{2}}$ is the dimensionality of the computational subspace. In the fidelity formula presented above, both unitarity and distance from the target operation are encapsulated. Note that $U_\mathrm{final}$ is the achieved unitary matrix after projection to the computational subspace and virtual phase compensation.

\section{Reinforcement Learning Modeling}

Formally, a reinforcement learning (RL) model can be described as a Markov Decision Process (MDP) tuple $(S, A, R, P, \gamma)$, where $S$ is the set of possible states, $A$ is the set of possible actions, $R$ is the distribution of reward given state action pairs $(s, a)$ where $s \in S$ and $a \in A$, $P$ is the transition probability distribution over next state $s_{t+1}$ given $(s, a)$, and $\gamma$ is the discount factor, a hyper-parameter to indicate the value of future rewards rather than the current reward. 

In the RL framework, a \emph{control policy} $\pi: S \rightarrow A$ defines what action to take at each state. In general RL models, the learning \emph{agent} interacts with an \emph{environment} by sending \emph{action} $a_t$ at time $t$ to an environment which is in some state $s_t$. The agent receives an \emph{observation}, \textit{i.e.} the next state of the environment $s_{t+1}$, and the \emph{reward} associated with the state and action $(s_t, a_t)$ \cite{reinforcement1996}. In this context, the control policy denoted by $\pi$ dictates which actions to take at each state. The goal is for the learning agent to find the optimal sequence of actions that will result in the maximum accumulated reward from the environment. In the following, we describe how the quantum gate design problem can be formulated in the RL model.

\subsection{Environment}

We can model the quantum gate design problem within the reinforcement learning framework such that a quantum simulator can serve as the environment. The objective is to learn the optimal control pulses to achieve a target unitary operation with maximum fidelity and lowest gate duration. We consider piecewise constant control pulses that can vary in 1 ns time steps. In this model, at each learning step, the control parameters of the system Hamiltonian are changed based on the received actions from the learning agent. Each learning episode consists of several learning steps that lead to the target unitary gate with a predefined fidelity or ends when the maximum number of steps is reached. We developed the quantum simulator environment based on the OpenAI interface \cite{brockman2016openai} \cite{baselines}.

\subsection{Observation}

The time evolution of this Hamiltonian causes a transition from the current quantum state to the next quantum state, \textit{i.e.} $\ket{\Psi(t+1)} = U \ket{\Psi(t)}$, where $U$ is the unitary operation during the learning step. One can model the gate design problem in the RL framework by returning the next quantum state as the observation. Here we propagate the unitary operation instead of the quantum states, such that at each learning step the unitary matrix $U_{t+1} = U_t U_{t-1} U_{t-2} ... U_{0}$ is calculated, where $U_{0}$ is the identity matrix. Therefore, we return the resultant unitary matrix $U_{t+1}$ as the observation ($S$ in the MDP tuple) to the learning agent.

\subsection{Action}

The action space implementation in an environment directly affects the choice and the performance of the RL learning algorithm. One may consider a discrete or a continuous action space for a physical environment \cite{lillicrap2015continuous} \cite{SuttonBook} \cite{LapanBook}.

In the discrete action space, at each time step, the learning agent only chooses the action from a discrete set provided by the environment. In the case of the quantum dot simulator environment, the direction of changing the control pulse at each time step can be encoded to an integer. For example, the numbers $\{0, 1, 2\}$ could respectively mean \{no change in tunnel coupling, increase the tunnel coupling by one step-size, decrease the tunnel coupling by one step-size\}. Here, the step-size is a predefined value to change the control parameter, which we refer to as the control step-size, to distinguish it from the RL algorithm updating step-size $\alpha$.

In the continuous action space, the action is a set of real numbers typically chosen from a probability distribution. For a quantum dot simulator, at each time step, the action set represents the value of the on-site energy of each dot and the tunnel coupling between the dots. The choice of these real values are constrained within a realistic range based on the physical system.

\subsection{Reward}

Since our learning objective is to design high-fidelity quantum gates with short gate duration, it is important to consider the gate duration and fidelity in the reward function. Here we considered a reward of $-1$ per learning step, so that the longer duration gates are more expensive. Experiments showed that the RL algorithms have higher performance when the fidelity of each learning step at non-terminal states is not considered in the reward function. However, if the learning episode resulted in the target unitary gate with the predefined fidelity, \textit{i.e.} we reached the terminal state, then the gate fidelity is magnified as a large positive reward. Setting the target fidelity as high as $(>0.999)$ caused most of the episodes to reach the maximum number of steps (200) without success and the learning process was slow. For faster convergence of the RL algorithm, we defined the terminal state as the state with a fidelity $>0.99$, however; if the fidelity $>0.999$, an even larger positive reward was given. We also considered a reward of $-1$ when the action resulted in a control parameter reaching the predefined realistic boundary.  

\section{Designing Multi-qubit Gates in a Discrete Action Space \label{CZ-DA}}

\subsection{Temporal Difference learning \label{TD}}

Temporal difference (TD) learning methods are powerful reinforcement learning algorithms that combine the advantages from Monte Carlo simulation and Dynamic Programming \cite{SuttonBook}. The TD methods are model-free which means we don't require complete model dynamics. Moreover TD methods are online, so that the algorithm learns at each step and it does not require us to wait until the end of the episode to learn. Here we utilize two of the most popular TD based RL algorithms, Q-learning and SARSA \cite{SuttonBook}, to design a \CZ gate for a quantum-dot simulator environment with a discrete action space.  

\begin{figure}
\includegraphics[width=\linewidth]{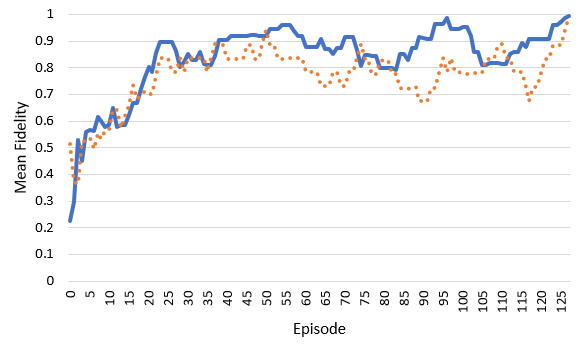}
\noindent \caption{\label{fig_td_fidelities}Mean fidelity vs. number of episodes for learning the \CZ gate using deep Q-Learning (orange dotted line) and deep SARSA (blue solid line). Note that the mean fidelity is calculated in groups of 10 episodes. In this experiment, the algorithm is terminated when the mean gate fidelity (over the last 10 episodes) reaches $>0.99$.}
\end{figure}

\subsection{Q-learning and SARSA \label{DQL_SARSA}}

In both algorithms, we considered a sequential neural network (NN) as a function approximator to learn the state-action value function $Q(s, a)$ which quantifies the action value at a given state $s$. For a two-dot system, there are 27 possible actions that correspond to permutations of three possible control pulses: the on-site energy of dot 0, the on-site energy of dot 1, and the tunnel coupling between the two dots. In both algorithm implementations, we used an $\epsilon$-greedy control policy derived from $Q(s, a)$. 

The $\epsilon$-greedy policy selects an action based on the state action value derived by the NN or takes a random action sampled from the action space. The probability of which method it uses to select an action is characterized by $\epsilon$. When the initial value of $\epsilon$ is high, the random sampling has a higher probability of action selection, and as the algorithm progresses in time, based on the  $\epsilon$ decay rate, the NN has a higher probability of predicting the action. 

In both algorithms, after each learning step, the temporal difference between the value of the (next-state, next-action) pair $Q(s', a')$, and the value of the (state, action) pair $Q(s, a)$ is calculated. Note that here, the $Q$ function is approximated through a sequential neural network (NN) with two hidden layers of size 64. The state which is the input of NN is the flattened unitary matrix observed from the dot-simulator, and the output of the NN is 27 predicted values associated with actions from the discrete action space. 

In deep Q-learning, the next action $a'$, given the next-state $s'$, is chosen as the one with highest predicted value from the NN (off-policy method). In deep SARSA, the next action $a'$, given the next-state $s'$, is chosen using the control policy (in our case $\epsilon$-greedy) and then the value of the (next-state, next-action) pair $Q(s', a')$ is obtained from the NN (on-policy method). The NN architecture for Q function estimation consist of two hidden layers of size 64 with "$\tanh$" activation functions and linear activation for the output layer. We used the Adam Optimizer \cite{kingma2014adam} with mean squared error loss function, learning rate of 0.001, and learning decay rate of 0.01. The following initial parameters were considered for the TD algorithms: step-size $\alpha=0.1$, discount factor $\gamma=0.9$, and $\epsilon$ decay multiplier $0.995$. 

In our simulations, the system parameters and constraints are chosen based on an experimental realization of a \CZ gate in a quantum-dot system \cite{Watson2018}. The biases on dots 0 and 1 are constrained in the range $[-750, 750]$ GHz and are initialized to 170 and 70 GHz, respectively. The tunnel coupling parameter is initialized to 2.5 GHz and constrained in the range $[0, 5]$ GHz. The Coulomb repulsion energies and Zeeman energies of the quantum dots are considered to be constant. The Coulomb repulsion energies for both dots were considered to be 845.2 GHz, the resonance frequencies of qubit 0 and qubit 1 were 18.4 and 19.7 GHz, respectively. 

\subsection{Results}

\begin{figure}
\begin{subfigure}{0.45\textwidth}
\includegraphics[width=\linewidth]{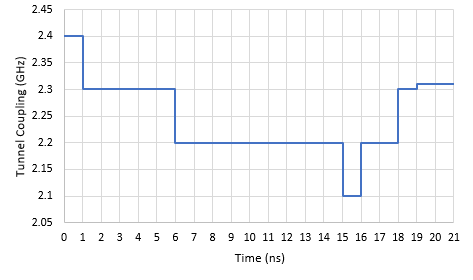}
\subcaption{}
\end{subfigure}
\begin{subfigure}{0.45\textwidth}
\includegraphics[width=\linewidth]{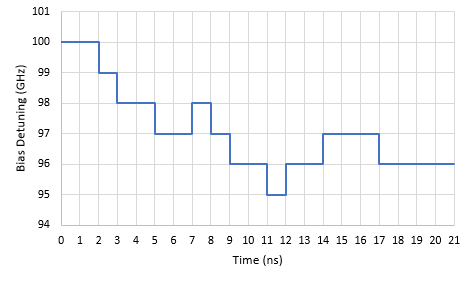}
\subcaption{}
\end{subfigure}
\noindent \caption{\label{fig-pulseDQL}The required tunnel coupling variations and on-site energy detuning to realize a high-fidelity \CZ gate. These results are achieved from running the Deep Q-Learning algorithm.  ~(a) Tunnel coupling variations  ~(b) Bias detuning between the two dots}
\end{figure}

Experiments showed that when the gate fidelity was not incorporated in the reward function at each learning step, including the gate fidelity in addition to the unitary matrix as the input state of the NN caused both algorithms to converge much more rapidly. Both deep SARSA and deep Q-Learning algorithms had similar performance as seen in \cref{fig_td_fidelities}. Using both algorithms, we could achieve the required control parameter variations to realize a $>0.999$ fidelity \CZ gate with the shortest gate duration of 21 ns. \Cref{fig-pulseDQL} shows the required tunnel coupling pulse and bias detuning pulse between the two dots.  

The discrete action space used to explore the quantum system control space introduces two main issues:
\begin{enumerate}
\item As depicted in \cref{fig-pulseDQL}, the control pulses only have small variations per time step. This is because in the discrete action space, the pulses can change only by a predefined control step-size whose value depends on the realistic experimental constraints. Although this is in our favor as we have full control over the rise/fall times of the piecewise-constant control pulses, there is the possibility of skipping the optimal solution because of the large predefined control step-size. To minimize this issue, we used an adaptive control step-size that is initially set to 1 GHz, but when the fidelity reaches as high as $>0.99$ or $>0.999$, the control step-size is adjusted to smaller values of 0.1 or 0.01 GHz, respectively. 

\item The discrete action space explores the large control landscape of a quantum system by small variations in control parameters at each learning step, so it can take a long time to converge to a good solution. Therefore, we initialized the control parameter with an educated guess derived from Ref.\cite{Watson2018}. 
\end{enumerate}

\section{Designing Multi-qubit gates in a Continuous Action Space }

\subsection{Policy Learning}

Modeling the quantum system as an environment with a continuous action space eliminates the control step-size issues by letting the RL algorithm assign real values to the control parameters at each learning step. For a continuous action space, the real valued actions are chosen from a probability distribution such as a normal distribution. The learning algorithm learns the control policy parameterized by the mean and standard deviation of a probability density function. 

Though there are many policy gradient algorithms, the Proximal Policy Optimization (PPO) \cite{schulman2017proximal} algorithm fits best to our specific control problem. First, in the PPO algorithm, the real-valued actions are sampled based on the same policy (probability distribution) at each episode; ensuring the rise/fall times of the control pulses would still remain realistic. Second, PPO avoids excessively large policy updates by a clipping mechanism \cite{schulman2017proximal}. Third, PPO has been proven to have good performance and reliability while being less computationally expensive than other algorithms in the continuous domain \cite{schulman2017proximal}. Finally, PPO is easily implementable in parallel using high performance computational resources. 
\begin{figure}
\begin{subfigure}{0.45\textwidth}
\includegraphics[width=\linewidth]{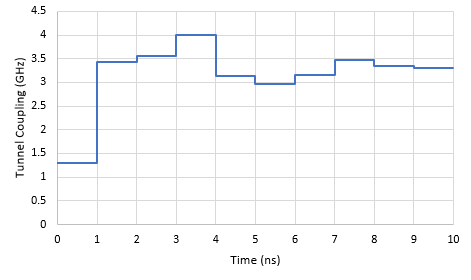}
\subcaption{}
\end{subfigure}
\begin{subfigure}{0.45\textwidth}
\includegraphics[width=\linewidth]{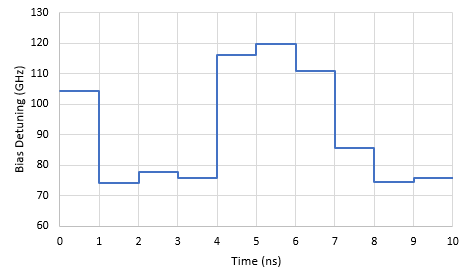}
\subcaption{}
\end{subfigure}
\noindent \caption{\label{fig-pulsePPO}The required tunnel coupling variations and on-site energy detuning to realize a high-fidelity \CZ gate. These results are achieved from running the Proximal Policy Optimization algorithm. ~(a) Tunnel coupling variations  ~(b) Bias detuning between the two dots}
\end{figure}
\subsection{Proximal Policy Optimization}

We utilize two sequential NNs named value-NN and policy-NN. Each has two hidden layers of size 64 with ``$\tanh$" activation functions. The value-NN is for state value estimation which quantifies what is the value of the given state. It takes the flattened unitary matrix as input and has one output for state value function estimation. The policy-NN is for the policy parameters estimation and takes the flattened unitary matrix as input and outputs the mean and standard deviation of a normal distribution. 

At each iteration of the PPO algorithm, a full control trajectory $(s_0, a_0, r_0, s_1, a_1, r_1, ..., s_T)$ for $T=200$ steps is constructed. At each step, the control policy from policy-NN is used to sample the real valued actions from the predefined ranges, $[-750, 750]$ GHz for biases, and $[0, 5]$ GHz for tunnel coupling. Then the advantage estimator of the full trajectory is computed utilizing the value-NN. Finally, we construct the loss function with respect to the policy parameters and optimize it using multiple steps of stochastic gradient ascent \cite{schulman2017proximal, baselines}. The initial parameters are considered as discounted factor of $\gamma=0.9$, and $\lambda=0.95$, clipping parameter  $\epsilon=0.2$, and learning rate of 0.001.

The PPO algorithm was originally designed to take advantage of parallel processing. At each iteration, several environments complete several control trajectories in parallel, while each environment samples the actions from the same probability distribution from policy-NN. After each iteration, the loss function is calculated utilizing the average advantage estimates of all of the parallel environments \cite{schulman2017proximal}. The parallel version of the PPO algorithm substantially speeds up the learning process and is suitable for designing gates for large qubit systems. In our simulations we utilized 40 cores to serve as the parallel environments.
\begin{figure}
\begin{subfigure}{0.45\textwidth}
\includegraphics[width=\linewidth]{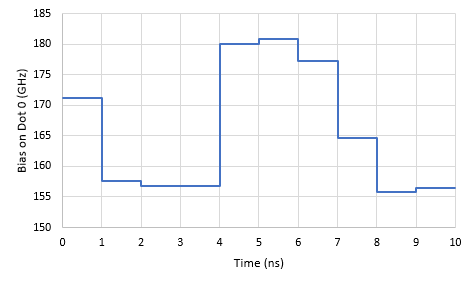}
\subcaption{}
\end{subfigure}
\begin{subfigure}{0.45\textwidth}
\includegraphics[width=\linewidth]{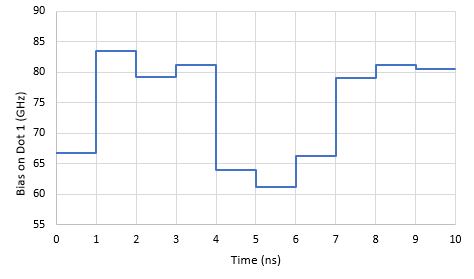}
\subcaption{}
\end{subfigure}
\noindent \caption{\label{fig-biasesPPO}The required on-site energy variations for each dot to achieve a high-fidelity \CZ gate. These results are achieved from running the Proximal Policy Optimization algorithm. ~(a) Bias on dot 0  ~(b) Bias on dot 1}
\end{figure}
\subsection{Results}

Using the PPO algorithm, we achieved a \CZ gate with $>0.999$ fidelity and duration of only 10 ns. The system parameters and constraints were initialized as described in \cref{DQL_SARSA}. The control pulses for tunnel coupling and bias detuning pulse between the two dots are shown in \cref{fig-pulsePPO}. As shown in \cref{fig-biasesPPO}, fast transitions exist in the resultant on-site energies (biases), however, they are consistent with realistic pulse profiles \cite{Medford2013c, Hensgens2018, Veldhorst2015, Watson2018}. 

\section{Conclusion}

The design, optimization, and control of quantum gates can largely benefit from machine-learning techniques. This is the first time that designing the multi-qubit gates for quantum-dot qubit systems is modeled within a deep reinforcement learning framework. Extensive simulations and experiments ensured a realistic and extensible quantum-dot simulator environment. The state-of-the-art RL algorithms based on Temporal Difference methods and Policy Gradient algorithms were investigated in both discrete and continuous action spaces, respectively. Our experiments showed that the quantum gates can efficiently be designed utilizing the parallel Proximal Policy Optimization algorithm. Using the discussed framework, we successfully achieved a $>0.999$ fidelity \CZ gate with duration of 10 ns for quantum-dot qubit chips. Moreover, the described RL modeling can be used to design multi-qubit gates for other quantum physical realizations.

\bibliographystyle{IEEEtrans}
\bibliography{bib}

\begin{thebibliography}{10}
\providecommand{\url}[1]{#1}
\csname url@samestyle\endcsname
\providecommand{\newblock}{\relax}
\providecommand{\bibinfo}[2]{#2}
\providecommand{\BIBentrySTDinterwordspacing}{\spaceskip=0pt\relax}
\providecommand{\BIBentryALTinterwordstretchfactor}{4}
\providecommand{\BIBentryALTinterwordspacing}{\spaceskip=\fontdimen2\font plus
\BIBentryALTinterwordstretchfactor\fontdimen3\font minus
  \fontdimen4\font\relax}
\providecommand{\BIBforeignlanguage}[2]{{%
\expandafter\ifx\csname l@#1\endcsname\relax
\typeout{** WARNING: IEEEtranS.bst: No hyphenation pattern has been}%
\typeout{** loaded for the language `#1'. Using the pattern for}%
\typeout{** the default language instead.}%
\else
\language=\csname l@#1\endcsname
\fi
#2}}
\providecommand{\BIBdecl}{\relax}
\BIBdecl

\bibitem{allen_2017}
J.~L. Allen, R.~Kosut, J.~Joo, P.~Leek, and E.~Ginossar, ``Optimal control of
  two qubits via a single cavity drive in circuit quantum electrodynamics,''
  \emph{Physical Review A}, vol.~95, no.~4, 2017.

\bibitem{brockman2016openai}
G.~Brockman, V.~Cheung, L.~Pettersson, J.~Schneider, J.~Schulman, J.~Tang, and
  W.~Zaremba, ``Openai gym,'' 2016.

\bibitem{Brunner2011}
\BIBentryALTinterwordspacing
R.~Brunner, Y.-S. Shin, T.~Obata, M.~Pioro-Ladri\`ere, T.~Kubo, K.~Yoshida,
  T.~Taniyama, Y.~Tokura, and S.~Tarucha, ``Two-qubit gate of combined
  single-spin rotation and interdot spin exchange in a double quantum dot,''
  \emph{Phys. Rev. Lett.}, vol. 107, p. 146801, Sep 2011. [Online]. Available:
  \url{https://link.aps.org/doi/10.1103/PhysRevLett.107.146801}
\BIBentrySTDinterwordspacing

\bibitem{Chunlin_Chen_2014}
\BIBentryALTinterwordspacing
C.~Chen, D.~Dong, H.-X. Li, J.~Chu, and T.-J. Tarn, ``Fidelity-based
  probabilistic q-learning for control of quantum systems,'' \emph{IEEE
  Transactions on Neural Networks and Learning Systems}, vol.~25, no.~5, p.
  920–933, May 2014. [Online]. Available:
  \url{http://dx.doi.org/10.1109/TNNLS.2013.2283574}
\BIBentrySTDinterwordspacing

\bibitem{daraeizadeh2019}
S.~Daraeizadeh, S.~P. Premaratne, X.~Song, M.~Perkowski, and A.~Y. Matsuura,
  ``Machine-learning based three-qubit gate for realization of a toffoli gate
  with cqed-based transmon systems,'' 2019.

\bibitem{baselines}
P.~Dhariwal, C.~Hesse, O.~Klimov, A.~Nichol, M.~Plappert, A.~Radford,
  J.~Schulman, S.~Sidor, Y.~Wu, and P.~Zhokhov, ``Openai baselines,''
  \url{https://github.com/openai/baselines}, 2017.

\bibitem{Foletti2009}
\BIBentryALTinterwordspacing
S.~Foletti, H.~Bluhm, D.~Mahalu, V.~Umansky, and A.~Yacoby, ``Universal quantum
  control of two-electron spin quantum bits using dynamic nuclear
  polarization,'' \emph{Nature Physics}, vol.~5, no.~12, pp. 903--908, Dec
  2009. [Online]. Available: \url{https://doi.org/10.1038/nphys1424}
\BIBentrySTDinterwordspacing

\bibitem{Fujisawa2004}
\BIBentryALTinterwordspacing
T.~Fujisawa, T.~Hayashi, H.~Cheong, Y.~Jeong, and Y.~Hirayama, ``Rotation and
  phase-shift operations for a charge qubit in a double quantum dot,''
  \emph{Physica E: Low-dimensional Systems and Nanostructures}, vol.~21, no.~2,
  pp. 1046 -- 1052, 2004, proceedings of the Eleventh International Conference
  on Modulated Semiconductor Structures. [Online]. Available:
  \url{http://www.sciencedirect.com/science/article/pii/S1386947703007513}
\BIBentrySTDinterwordspacing

\bibitem{Goerz_2017}
\BIBentryALTinterwordspacing
M.~H. Goerz, F.~Motzoi, K.~B. Whaley, and C.~P. Koch, ``Charting the circuit
  qed design landscape using optimal control theory,'' \emph{npj Quantum
  Information}, vol.~3, no.~1, Sep 2017. [Online]. Available:
  \url{http://dx.doi.org/10.1038/s41534-017-0036-0}
\BIBentrySTDinterwordspacing

\bibitem{Golovach2006}
\BIBentryALTinterwordspacing
V.~N. Golovach, M.~Borhani, and D.~Loss, ``Electric-dipole-induced spin
  resonance in quantum dots,'' \emph{Phys. Rev. B}, vol.~74, p. 165319, Oct
  2006. [Online]. Available:
  \url{https://link.aps.org/doi/10.1103/PhysRevB.74.165319}
\BIBentrySTDinterwordspacing

\bibitem{Yung_guzik14}
\BIBentryALTinterwordspacing
M.~h~Yung, J.~D. Whitfield, D.~G.~T. S.~Boixo, and A.~Aspuru-Guzik,
  ``Introduction to quantum algorithms for physics and chemistry,''
  \emph{Advances in Chemical Physics}, vol.~21, March 2014. [Online].
  Available: \url{https://doi.org/10.1002/9781118742631.ch03}
\BIBentrySTDinterwordspacing

\bibitem{Hanson2007}
\BIBentryALTinterwordspacing
R.~Hanson, L.~P. Kouwenhoven, J.~R. Petta, S.~Tarucha, and L.~M.~K.
  Vandersypen, ``Spins in few-electron quantum dots,'' \emph{Rev. Mod. Phys.},
  vol.~79, pp. 1217--1265, Oct 2007. [Online]. Available:
  \url{https://link.aps.org/doi/10.1103/RevModPhys.79.1217}
\BIBentrySTDinterwordspacing

\bibitem{Hayashi2003}
\BIBentryALTinterwordspacing
T.~Hayashi, T.~Fujisawa, H.~D. Cheong, Y.~H. Jeong, and Y.~Hirayama, ``Coherent
  manipulation of electronic states in a double quantum dot,'' \emph{Phys. Rev.
  Lett.}, vol.~91, p. 226804, Nov 2003. [Online]. Available:
  \url{https://link.aps.org/doi/10.1103/PhysRevLett.91.226804}
\BIBentrySTDinterwordspacing

\bibitem{Hensgens2018}
\BIBentryALTinterwordspacing
T.~Hensgens, ``{Emulating Fermi-Hubbard physics with quantum dots: from few to
  more and how to},'' 2018. [Online]. Available:
  \url{https://doi.org/10.4233/uuid:b71f3b0b-73a0-4996-896c-84ed43e72035}
\BIBentrySTDinterwordspacing

\bibitem{reinforcement1996}
L.~P. Kaelbling, M.~L. Littman, and A.~W. Moore, ``Reinforcement learning: A
  survey,'' 1996.

\bibitem{Kawakami2014}
\BIBentryALTinterwordspacing
E.~Kawakami, P.~Scarlino, D.~R. Ward, F.~R. Braakman, D.~E. Savage, M.~G.
  Lagally, M.~Friesen, S.~N. Coppersmith, M.~A. Eriksson, and L.~M.~K.
  Vandersypen, ``Electrical control of a long-lived spin qubit in a si/sige
  quantum dot,'' \emph{Nature Nanotechnology}, vol.~9, no.~9, pp. 666--670, Sep
  2014. [Online]. Available: \url{https://doi.org/10.1038/nnano.2014.153}
\BIBentrySTDinterwordspacing

\bibitem{Khaneja}
\BIBentryALTinterwordspacing
N.~Khaneja, T.~Reiss, C.~Kehlet, T.~Schulte-Herbrüggen, and S.~J. Glaser,
  ``Optimal control of coupled spin dynamics: design of nmr pulse sequences by
  gradient ascent algorithms,'' \emph{Journal of Magnetic Resonance}, vol. 172,
  no.~2, pp. 296 -- 305, 2005. [Online]. Available:
  \url{http://www.sciencedirect.com/science/article/pii/S1090780704003696}
\BIBentrySTDinterwordspacing

\bibitem{Kim2014}
\BIBentryALTinterwordspacing
D.~Kim, Z.~Shi, C.~B. Simmons, D.~R. Ward, J.~R. Prance, T.~S. Koh, J.~K.
  Gamble, D.~E. Savage, M.~G. Lagally, M.~Friesen, S.~N. Coppersmith, and M.~A.
  Eriksson, ``Quantum control and process tomography of a semiconductor quantum
  dot hybrid qubit,'' \emph{Nature}, vol. 511, no. 7507, pp. 70--74, Jul 2014.
  [Online]. Available: \url{https://doi.org/10.1038/nature13407}
\BIBentrySTDinterwordspacing

\bibitem{kingma2014adam}
D.~P. Kingma and J.~Ba, ``Adam: A method for stochastic optimization,'' 2014.

\bibitem{Koh2012}
\BIBentryALTinterwordspacing
T.~S. Koh, J.~K. Gamble, M.~Friesen, M.~A. Eriksson, and S.~N. Coppersmith,
  ``Pulse-gated quantum-dot hybrid qubit,'' \emph{Phys. Rev. Lett.}, vol. 109,
  p. 250503, Dec 2012. [Online]. Available:
  \url{https://link.aps.org/doi/10.1103/PhysRevLett.109.250503}
\BIBentrySTDinterwordspacing

\bibitem{Koppens2006}
\BIBentryALTinterwordspacing
F.~H.~L. Koppens, C.~Buizert, K.~J. Tielrooij, I.~T. Vink, K.~C. Nowack,
  T.~Meunier, L.~P. Kouwenhoven, and L.~M.~K. Vandersypen, ``Driven coherent
  oscillations of a single electron spin in a quantum dot,'' \emph{Nature},
  vol. 442, no. 7104, pp. 766--771, Aug 2006. [Online]. Available:
  \url{https://doi.org/10.1038/nature05065}
\BIBentrySTDinterwordspacing

\bibitem{Kuhlman2013}
A.~V. Kuhlman, J.~Houel, A.~Ludwig, L.~Greuter, D.~Reuter, A.~D. Wieck,
  M.~Poggio, and R.~J. Warburton, ``Charge noise and spin noise in a
  semiconductor quantum device,'' \emph{Nature Physics}, vol.~9, p. 570–575,
  2013.

\bibitem{LapanBook}
M.~Lapan, \emph{Deep Reinforcement Learning Hands-On}.\hskip 1em plus 0.5em
  minus 0.4em\relax Packt Publishing, 2018.

\bibitem{lillicrap2015continuous}
T.~P. Lillicrap, J.~J. Hunt, A.~Pritzel, N.~Heess, T.~Erez, Y.~Tassa,
  D.~Silver, and D.~Wierstra, ``Continuous control with deep reinforcement
  learning,'' 2015.

\bibitem{Loss1998}
\BIBentryALTinterwordspacing
D.~Loss and D.~P. DiVincenzo, ``Quantum computation with quantum dots,''
  \emph{Phys. Rev. A}, vol.~57, pp. 120--126, Jan 1998. [Online]. Available:
  \url{https://link.aps.org/doi/10.1103/PhysRevA.57.120}
\BIBentrySTDinterwordspacing

\bibitem{Machnes_2011}
\BIBentryALTinterwordspacing
S.~Machnes, U.~Sander, S.~J. Glaser, P.~de~Fouquières, A.~Gruslys,
  S.~Schirmer, and T.~Schulte-Herbrüggen, ``Comparing, optimizing, and
  benchmarking quantum-control algorithms in a unifying programming
  framework,'' \emph{Physical Review A}, vol.~84, no.~2, Aug 2011. [Online].
  Available: \url{http://dx.doi.org/10.1103/PhysRevA.84.022305}
\BIBentrySTDinterwordspacing

\bibitem{Martins2016}
\BIBentryALTinterwordspacing
F.~Martins, F.~K. Malinowski, P.~D. Nissen, E.~Barnes, S.~Fallahi, G.~C.
  Gardner, M.~J. Manfra, C.~M. Marcus, and F.~Kuemmeth, ``Noise suppression
  using symmetric exchange gates in spin qubits,'' \emph{Phys. Rev. Lett.},
  vol. 116, p. 116801, Mar 2016. [Online]. Available:
  \url{https://link.aps.org/doi/10.1103/PhysRevLett.116.116801}
\BIBentrySTDinterwordspacing

\bibitem{mcclean2017openfermion}
J.~R. McClean, K.~J. Sung, I.~D. Kivlichan, Y.~Cao, C.~Dai, E.~S. Fried,
  C.~Gidney, B.~Gimby, P.~Gokhale, T.~Häner, T.~Hardikar, V.~Havlíček,
  O.~Higgott, C.~Huang, J.~Izaac, Z.~Jiang, X.~Liu, S.~McArdle, M.~Neeley,
  T.~O'Brien, B.~O'Gorman, I.~Ozfidan, M.~D. Radin, J.~Romero, N.~Rubin,
  N.~P.~D. Sawaya, K.~Setia, S.~Sim, D.~S. Steiger, M.~Steudtner, Q.~Sun,
  W.~Sun, D.~Wang, F.~Zhang, and R.~Babbush, ``Openfermion: The electronic
  structure package for quantum computers,'' 2017.

\bibitem{Medford2013b}
\BIBentryALTinterwordspacing
J.~Medford, J.~Beil, J.~M. Taylor, S.~D. Bartlett, A.~C. Doherty, E.~I. Rashba,
  D.~P. DiVincenzo, H.~Lu, A.~C. Gossard, and C.~M. Marcus, ``Self-consistent
  measurement and state tomography of an exchange-only spin qubit,''
  \emph{Nature Nanotechnology}, vol.~8, no.~9, pp. 654--659, Sep 2013.
  [Online]. Available: \url{https://doi.org/10.1038/nnano.2013.168}
\BIBentrySTDinterwordspacing

\bibitem{Medford2013a}
\BIBentryALTinterwordspacing
J.~Medford, J.~Beil, J.~M. Taylor, E.~I. Rashba, H.~Lu, A.~C. Gossard, and
  C.~M. Marcus, ``Quantum-dot-based resonant exchange qubit,'' \emph{Phys. Rev.
  Lett.}, vol. 111, p. 050501, Jul 2013. [Online]. Available:
  \url{https://link.aps.org/doi/10.1103/PhysRevLett.111.050501}
\BIBentrySTDinterwordspacing

\bibitem{Medford2013c}
\BIBentryALTinterwordspacing
J.~R. Medford, ``{Spin Qubits in Double and Triple Quantum Dots},'' 2013.
  [Online]. Available:
  \url{http://nrs.harvard.edu/urn-3:HUL.InstRepos:11156788}
\BIBentrySTDinterwordspacing

\bibitem{google_trpo}
V.~N. S. . H.~N. Murphy Yuezhen~Niu, Sergio~Boixo, ``Universal quantum control
  through deep reinforcement learning,'' \emph{npj Quantum Inf}, vol.~5,
  no.~33, 2019.

\bibitem{Nowack2007}
\BIBentryALTinterwordspacing
K.~C. Nowack, F.~H.~L. Koppens, Y.~V. Nazarov, and L.~M.~K. Vandersypen,
  ``Coherent control of a single electron spin with electric fields,''
  \emph{Science}, vol. 318, no. 5855, pp. 1430--1433, 2007. [Online].
  Available: \url{https://science.sciencemag.org/content/318/5855/1430}
\BIBentrySTDinterwordspacing

\bibitem{Pedersen_2007}
\BIBentryALTinterwordspacing
L.~H. Pedersen, N.~M. Møller, and K.~Mølmer, ``Fidelity of quantum
  operations,'' \emph{Physics Letters A}, vol. 367, no. 1-2, p. 47–51, Jul
  2007. [Online]. Available:
  \url{http://dx.doi.org/10.1016/j.physleta.2007.02.069}
\BIBentrySTDinterwordspacing

\bibitem{Petersson2010}
\BIBentryALTinterwordspacing
K.~D. Petersson, J.~R. Petta, H.~Lu, and A.~C. Gossard, ``Quantum coherence in
  a one-electron semiconductor charge qubit,'' \emph{Phys. Rev. Lett.}, vol.
  105, p. 246804, Dec 2010. [Online]. Available:
  \url{https://link.aps.org/doi/10.1103/PhysRevLett.105.246804}
\BIBentrySTDinterwordspacing

\bibitem{Petit2020}
L.~Petit, H.~Eenink, W.~Lawrie, N.~Hendrickx., S.~Philips, J.~Clarke,
  L.~Vandersypen, and M.~Veldhorst, ``Universal quantum logic in hot silicon
  qubits,'' \emph{Nature}, vol. 580, p. 355–359, 2020.

\bibitem{Petta2005}
\BIBentryALTinterwordspacing
J.~R. Petta, A.~C. Johnson, J.~M. Taylor, E.~A. Laird, A.~Yacoby, M.~D. Lukin,
  C.~M. Marcus, M.~P. Hanson, and A.~C. Gossard, ``Coherent manipulation of
  coupled electron spins in semiconductor quantum dots,'' \emph{Science}, vol.
  309, no. 5744, pp. 2180--2184, 2005. [Online]. Available:
  \url{https://science.sciencemag.org/content/309/5744/2180}
\BIBentrySTDinterwordspacing

\bibitem{Rach_2015}
\BIBentryALTinterwordspacing
N.~Rach, M.~M. Müller, T.~Calarco, and S.~Montangero, ``Dressing the
  chopped-random-basis optimization: A bandwidth-limited access to the
  trap-free landscape,'' \emph{Physical Review A}, vol.~92, no.~6, Dec 2015.
  [Online]. Available: \url{http://dx.doi.org/10.1103/PhysRevA.92.062343}
\BIBentrySTDinterwordspacing

\bibitem{schulman2017proximal}
J.~Schulman, F.~Wolski, P.~Dhariwal, A.~Radford, and O.~Klimov, ``Proximal
  policy optimization algorithms,'' 2017.

\bibitem{Shi2012}
\BIBentryALTinterwordspacing
Z.~Shi, C.~B. Simmons, J.~R. Prance, J.~K. Gamble, T.~S. Koh, Y.-P. Shim,
  X.~Hu, D.~E. Savage, M.~G. Lagally, M.~A. Eriksson, M.~Friesen, and S.~N.
  Coppersmith, ``Fast hybrid silicon double-quantum-dot qubit,'' \emph{Phys.
  Rev. Lett.}, vol. 108, p. 140503, Apr 2012. [Online]. Available:
  \url{https://link.aps.org/doi/10.1103/PhysRevLett.108.140503}
\BIBentrySTDinterwordspacing

\bibitem{Shim2016}
\BIBentryALTinterwordspacing
Y.-P. Shim and C.~Tahan, ``Charge-noise-insensitive gate operations for
  always-on, exchange-only qubits,'' \emph{Physical Review B}, vol.~93, no.~12,
  Mar 2016. [Online]. Available:
  \url{http://dx.doi.org/10.1103/PhysRevB.93.121410}
\BIBentrySTDinterwordspacing

\bibitem{Spiteri_2018}
R.~J. Spiteri, M.~Schmidt, J.~Ghosh, E.~Zahedinejad, and B.~C. Sanders,
  ``Quantum control for high-fidelity multi-qubit gates,'' \emph{New Journal of
  Physics}, vol.~20, no.~11, p. 113009, nov 2018.

\bibitem{Steck2020}
\BIBentryALTinterwordspacing
D.~A. Steck. (2020, Apr.) {Quantum and Atom Optics}. Revision 0.13.1. Accessed
  05/01/2020. [Online]. Available: \url{http://steck.us/teaching}
\BIBentrySTDinterwordspacing

\bibitem{SuttonBook}
R.~S. Sutton and A.~G. Barto, \emph{Reinforcement Learning: An
  Introduction}.\hskip 1em plus 0.5em minus 0.4em\relax Cambridge,
  Massachusetts ; London, England : The MIT Press, 2018.

\bibitem{Veldhorst2015}
\BIBentryALTinterwordspacing
M.~Veldhorst, C.~H. Yang, J.~C.~C. Hwang, W.~Huang, J.~P. Dehollain, J.~T.
  Muhonen, S.~Simmons, A.~Laucht, F.~E. Hudson, K.~M. Itoh, A.~Morello, and
  A.~S. Dzurak, ``A two-qubit logic gate in silicon,'' \emph{Nature}, vol. 526,
  no. 7573, pp. 410--414, Oct 2015. [Online]. Available:
  \url{https://doi.org/10.1038/nature15263}
\BIBentrySTDinterwordspacing

\bibitem{Watson2018}
\BIBentryALTinterwordspacing
T.~F. Watson, S.~G.~J. Philips, E.~Kawakami, D.~R. Ward, P.~Scarlino,
  M.~Veldhorst, D.~E. Savage, M.~G. Lagally, M.~Friesen, S.~N. Coppersmith,
  M.~A. Eriksson, and L.~M.~K. Vandersypen, ``A programmable two-qubit quantum
  processor in silicon,'' \emph{Nature}, vol. 555, no. 7698, pp. 633--637, Mar
  2018. [Online]. Available: \url{https://doi.org/10.1038/nature25766}
\BIBentrySTDinterwordspacing

\bibitem{Yang2011}
\BIBentryALTinterwordspacing
S.~Yang, X.~Wang, and S.~Das~Sarma, ``Generic hubbard model description of
  semiconductor quantum-dot spin qubits,'' \emph{Phys. Rev. B}, vol.~83, p.
  161301, Apr 2011. [Online]. Available:
  \url{https://link.aps.org/doi/10.1103/PhysRevB.83.161301}
\BIBentrySTDinterwordspacing

\bibitem{Zahedinejad_2015}
\BIBentryALTinterwordspacing
E.~Zahedinejad, J.~Ghosh, and B.~C. Sanders, ``High-fidelity single-shot
  toffoli gate via quantum control,'' \emph{Physical Review Letters}, vol. 114,
  no.~20, May 2015. [Online]. Available:
  \url{http://dx.doi.org/10.1103/PhysRevLett.114.200502}
\BIBentrySTDinterwordspacing

\bibitem{Zahedinejad_2016}
\BIBentryALTinterwordspacing
------, ``Designing high-fidelity single-shot three-qubit gates: A
  machine-learning approach,'' \emph{Physical Review Applied}, vol.~6, no.~5,
  Nov 2016. [Online]. Available:
  \url{http://dx.doi.org/10.1103/PhysRevApplied.6.054005}
\BIBentrySTDinterwordspacing

\bibitem{Zwaneburg2013}
\BIBentryALTinterwordspacing
F.~A. Zwanenburg, A.~S. Dzurak, A.~Morello, M.~Y. Simmons, L.~C.~L. Hollenberg,
  G.~Klimeck, S.~Rogge, S.~N. Coppersmith, and M.~A. Eriksson, ``Silicon
  quantum electronics,'' \emph{Rev. Mod. Phys.}, vol.~85, pp. 961--1019, Jul
  2013. [Online]. Available:
  \url{https://link.aps.org/doi/10.1103/RevModPhys.85.961}
\BIBentrySTDinterwordspacing

\end{thebibliography}

\end{document}